\documentclass[structabstract]{aa}  
\usepackage{graphicx}
\usepackage{txfonts}
\usepackage{natbib}
\begin{document}
\title{Hierarchical fragmentation and collapse signatures in a high-mass starless region\thanks{Based on
    observations carried out with the IRAM Plateau de Bure
    Interferometer. IRAM is supported by INSU/CNRS (France), MPG
    (Germany) and IGN (Spain). The data are available in electronic
    form at the CDS via anonymous ftp to cdsarc.u-strasbg.fr
    (130.79.128.5) or via
    http://cdsweb.u-strasbg.fr/cgi-bin/qcat?J/A+A/}.}


   \author{H.~Beuther
          \inst{1}
          \and
          Th.~Henning
          \inst{1}
          \and
          H.~Linz
          \inst{1}
          \and
          S.~Feng
          \inst{1}
          \and
          S.E.~Ragan
          \inst{2}
          \and
          R.J.~Smith
          \inst{3}
          \and
          S.~Bihr
          \inst{1}
          \and
          T.~Sakai
          \inst{4}
          \and
          R.~Kuiper
          \inst{5}
           }
   \institute{$^1$ Max Planck Institute for Astronomy, K\"onigstuhl 17,
              69117 Heidelberg, Germany, \email{name@mpia.de}\\
              $^2$ University of Leeds, Leeds, LS2 9JT, UK\\
              $^3$ Jodrell Bank Centre for Astrophysics, School of Physics and Astronomy, The University of Manchester, Oxford Road, Manchester, M13 9PL, U.K.\\
              $^4$ Graduate School of Informatics and Engineering, The University of Electro-Communications, Chofu, 182-8585, Tokyo, Japan\\
              $^5$ Institute of Astronomy and Astrophysics, University of T\"ubingen, Auf der Morgenstelle 10, 72076 T\"ubingen, Germany
}

   \date{Version of \today}

\abstract
 {} 
   {Understanding the fragmentation and collapse properties of the
     dense gas during the onset of high-mass star formation.}
   {We observed the massive ($\sim$800\,M$_{\odot}$) starless gas
     clump IRDC\,18310-4 with the Plateau de Bure Interferometer
     (PdBI) at sub-arcsecond resolution in the 1.07\,mm continuum and
     N$_2$H$^+$(3--2) line emission.}
{Zooming from a
     single-dish low-resolution map to previous 3\,mm PdBI data, and
     now the new 1.07\,mm continuum observations, the sub-structures
     hierarchically fragment on the increasingly smaller spatial
     scales.  While the fragment separations may still be roughly
     consistent with pure thermal Jeans fragmentation, the derived
     core masses are almost two orders of magnitude larger than the
     typical Jeans mass at the given densities and
     temperatures. However, the data can be reconciled with models
     using non-homogeneous initial density structures, turbulence
     and/or magnetic fields. While most sub-cores remain
     (far-)infrared dark even at 70\,$\mu$m, we identify weak
     70\,$\mu$m emission toward one core with a comparably low
     luminosity of $\sim$16\,L$_{\odot}$, re-enforcing the general
     youth of the region. The spectral line data always exhibit
     multiple spectral components toward each core with comparably
     small line widths for the individual components (in the 0.3 to
     1.0\,km\,s$^{-1}$ regime). Based on single-dish C$^{18}$O(2--1)
     data we estimate a low virial-to-gas-mass ratio $\leq 0.25$. We
     discuss that the likely origin of these spectral properties may
     be the global collapse of the original gas clump that results in
     multiple spectral components along each line of sight.  Even
     within this dynamic picture the individual collapsing gas cores
     appear to have very low levels of internal turbulence.}
{}

\titlerunning{Hierarchical fragmentation and collapse signatures}

\maketitle

\section{Introduction}
\label{intro}

Independent of the various formation scenarios for high-mass stars
that are discussed extensively in the literature (e.g.,
\citealt{zinnecker2007,beuther2006b,tan2014}), the initial conditions
required to allow high-mass star formation at all are still poorly
characterized. The initial debate even ranged around the question
whether high-mass starless gas clumps should exist at all, or whether
the collapse of massive gas clumps starts immediately without any
clear starless phase in the high-mass regime (e.g.,
\citealt{motte2007}). Recent studies indicate that the time span
during which massive gas clumps exist without embedded star formation
is relatively short (on the order of 50000\,yrs), but nevertheless,
high-mass starless gas clumps do exist (e.g.,
\citealt{russeil2010,tackenberg2012,csengeri2014}). Current questions
in that field are: are high-mass gas clumps dominated by a single
fragment or do we witness strong fragmentation during earliest
evolutionary stages (e.g., \citealt{bontemps2010,zhang2015})? What are
the kinematic properties of the gas \citep{dobbs2014}? Are the clumps
sub- or super-virial \citep{tan2014}?  Do we see streaming motions
indicative of turbulent flows (e.g.,
\citealt{bergin2004,vazquez2006,heitsch2008,banerjee2009,motte2014,dobbs2014})?

To address such questions, one needs to investigate the earliest
evolutionary stages prior to the existence of embedded heating and
outflow sources that could quickly destroy any signatures of the early
kinematic and fragmentation properties. Furthermore, high-spatial
resolution is mandatory to resolve the important sub-structures at
typical distances of several kpc. To get a taste of the important
scales, we can estimate typical Jeans lengths for high-mass
star-forming regions: for example, average densities $\rho$ of
high-mass star-forming gas clumps on $\sim 0.5$\,pc scale in the
$10^5$\,cm$^{-3}$ density regime at low temperatures of 15\,K result
in typical Jeans fragmentation scales of $\sim$10000\,AU. Going to
smaller spatial scales ($\leq 0.1$\,pc), the embedded cores have
higher densities in the $10^6$\,cm$^{-3}$ regime that result in much
smaller Jeans fragmentation scales on the order of 4000\,AU (e.g.,
\citealt{beuther2013a}).

Our target region IRDC\,18310-4 is a 70\,$\mu$m dark high-mass
starless region at 4.9\,kpc distance that was previously observed with
the PdBI in the CD configuration in the 3\,mm continuum and the
N$_2$H$^+$(1--0) emission with a spatial resolution of $4.3''\times
3.0''$ \citep{beuther2013a}.  We identified fragmentation and core
formation of the still starless gas clump, and approximate separations
between the separate cores are on the order of 25000\,AU.
Fragmentation of these cores could not be resolved by the previous
observations.  The simultaneously observed N$_2$H$^+$(1--0) data
exhibit at least two velocity components.  Such
multi-velocity-components are the expected signatures of collapsing
and fragmenting gas-clumps \citep{smith2013}.

In a hierarchically structured star-forming region, the cores are the
supposed entities where further fragmentation should take place and
which likely contain bound multiple systems. Studying this region now
at 1.07\,mm in the dust continuum and N$_2$H$^+(3-2)$ at sub-arcsecond
resolution reveals the hierarchical fragmentation and kinematic
properties of the densest cores at the onset of high-mass star
formation.

\section{Observations} 
\label{obs}

\subsection{Plateau de Bure Interferometer}

The target region IRDC\,18310-4 was observed with the Plateau de Bure
Interferometer at 1.07\,mm wavelength in the B and C configuration in
4 tracks in March and November 2013. The projected baselines extended
to approximately 420\,m. The 1\,mm receivers were tuned to
279.512\,GHz in the lower sideband. At the given wavelength, the FWHM
of the primary beam is approximately $18''$, and we used a small
2-field mosaic to cover our region of interest. Phase and amplitude
calibration was conducted with regular observations of the quasars
2013+370 and 1749+096. The absolute flux level was calibrated with
MWC349 and bandpass calibration was done with 3C84. We estimate the
final flux accuracy to be correct to within $\sim 15\%$. The phase
reference center is RA (J2000.0) 18:33:39.532 and Dec (J2000.0)
-08:21:09.60, and the velocity of rest $v_{\rm{lsr}}$ is
$\sim$86.5\,km\,s$^{-1}$. While the channel spacing of the correlator
was 0.084\,km\,s$^{-1}$, to increase the signal-to-noise ratio we
imaged the data with 0.2\,km\,s$^{-1}$ spectral resolution. For the
continuum and line data we used different weighting systems between
natural and uniform weighting to improve the signal-to-noise ratio in
particular for the line data. The resulting synthesized beam for the
continuum and N$_2$H$^+$(3--2) data were $0.6''\times0.49''$ (with a
position angle of $11^o$) and $1.13''\times0.5''$ (with a position
angle of $13^o$), respectively.  The corresponding $1\sigma$ rms for
the two data sets is $\sim$0.6\,mJy\,beam$^{-1}$ and
$\sim$15\,mJy\,beam$^{-1}$ measured in a line-free channel of
0.2\,km\,s$^{-1}$ for the line data.

\subsection{Herschel PACS}
\label{Subsec:Herschel-obs}

The Herschel data were already presented in \citet{beuther2013b}.
However, they were fully re-calibrated for this analysis with
particular emphasis to applying the most recent pointing and
astrometry solutions. The region containing IRDC\,18310-4 was observed
with the Herschel spacecraft \citep{A&ASpecialIssue-HERSCHEL} within
the Key Project EPoS \citep{ragan2012b}.  The related observations
utilizing the bolometer cameras of PACS \citep{A&ASpecialIssue-PACS}
took place on April 19, 2011 (operational day 705) and were contained
in the observational IDs 1342219060--63. The PACS prime mode was
employed with a nominal scanning velocity of
20$''$\,second$^{-1}$. Data were taken in all three PACS filters (70,
100, and 160\,$\mu$m) and resulted in maps with roughly 10 arcmin
field-of-view. We re-reduced the existing Herschel archive data,
starting from the Level1 data. These we retrieved from the previous
bulk processing with the HIPE 12.1 software version, currently
contained in the HSA. We included the newly developed data reduction
step calcAttitude: a correction of the frames pointing product based
on the Herschel gyroscope house-keeping \citep{sanchez-portal2014}.
This often improves the absolute pointing accuracy and mitigates the
pointing jitter effect on individual frames. We used these corrected
frames and performed the individual detector pixel distortion
correction, the de-striping, the removal of 1/f noise, and the final
map projection using Scanamorphos \citep{roussel2013}, version 24,
employing the ``galactic'' option. The resulting full width half
maximum (FWHM) of the point spread function (PSF) at 70, 100, and
160\,$\mu$m are $5.6''$, $6.8''$ and $11.3''$, respectively.

\section{Results}

\subsection{Millimeter continuum emission}
\label{cont}

\begin{figure*}[htb]
\includegraphics[width=0.99\textwidth]{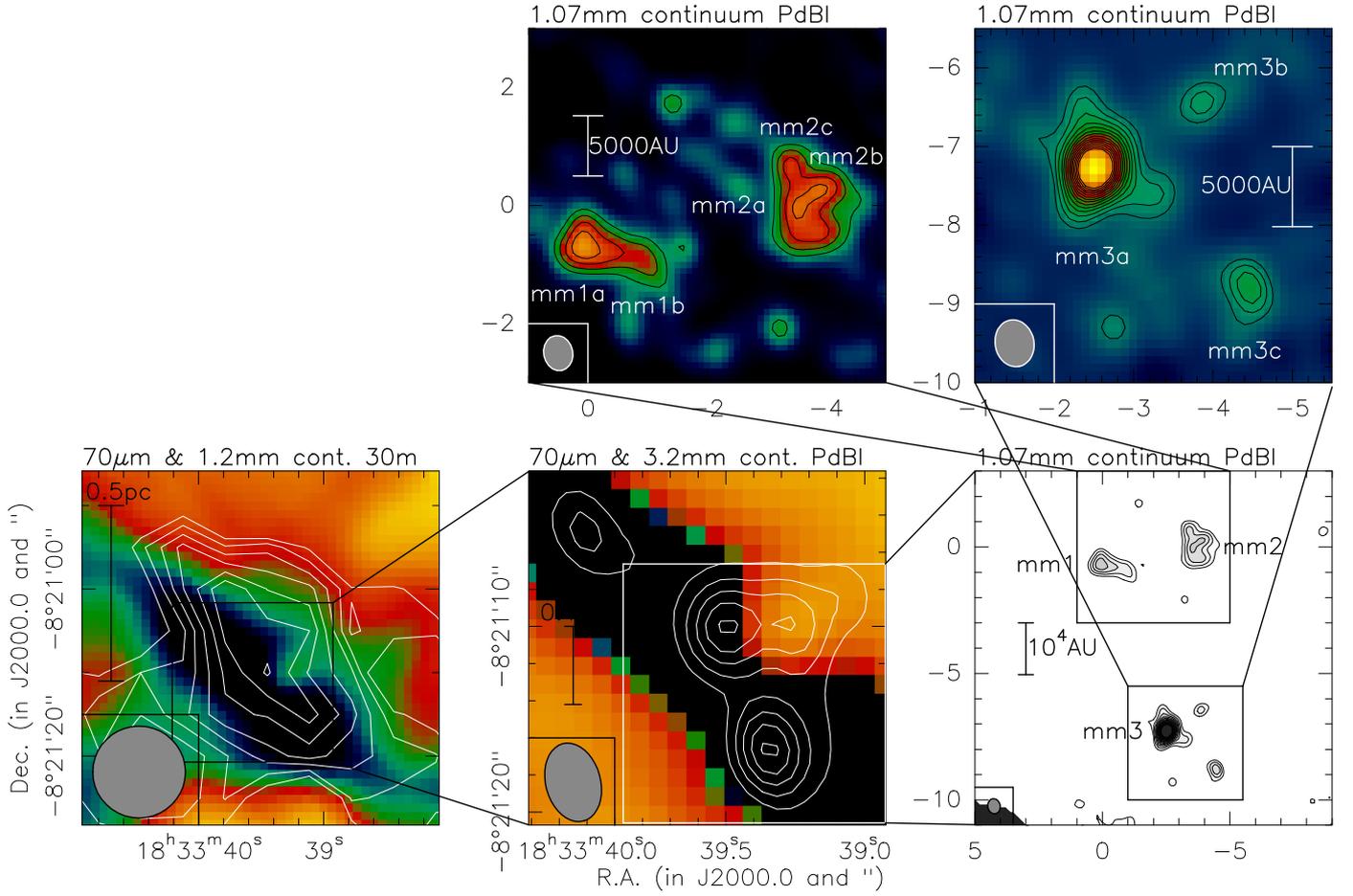}
\caption{Compilation of the continuum images in IRDC\,18310-4. The
  bottom-left and middle panels show in color the Herschel 70\,$\mu$m
  image with a stretch going dark for low values
  (\citealt{beuther2013a}, linear and log stretch for clarity,
  respectively). The white contours in the left panel show the 1.2\,mm
  MAMBO continuum observations starting from 4$\sigma$ and continuing
  in 1$\sigma$ steps with a $1\sigma$ value of 13\,mJy\,beam$^{-1}$.
  The contours in the bottom-middle panels presents the 3\,mm
  continuum data from \citet{beuther2013a} starting from 3$\sigma$ and
  continuing in 2$\sigma$ steps with a $1\sigma$ value of
  0.08\,mJy\,beam$^{-1}$. The bottom-right panel then shows the new
  1.07\,mm continuum observations starting from 3$\sigma$ and
  continuing in 1$\sigma$ steps with a $1\sigma$ value of
  0.6\,mJy\,beam$^{-1}$. The two top panels show zooms of the 1.07
  continuum data with the same contour levels but a different color
  stretch to highlight the sub-structures. Each panel presents
  scale-bars and the corresponding synthesized beams. The 70\,$\mu$m
  PSF is $5.6''$.}
\label{continuum}
\end{figure*}

\begin{figure*}[htb]
\includegraphics[width=0.99\textwidth]{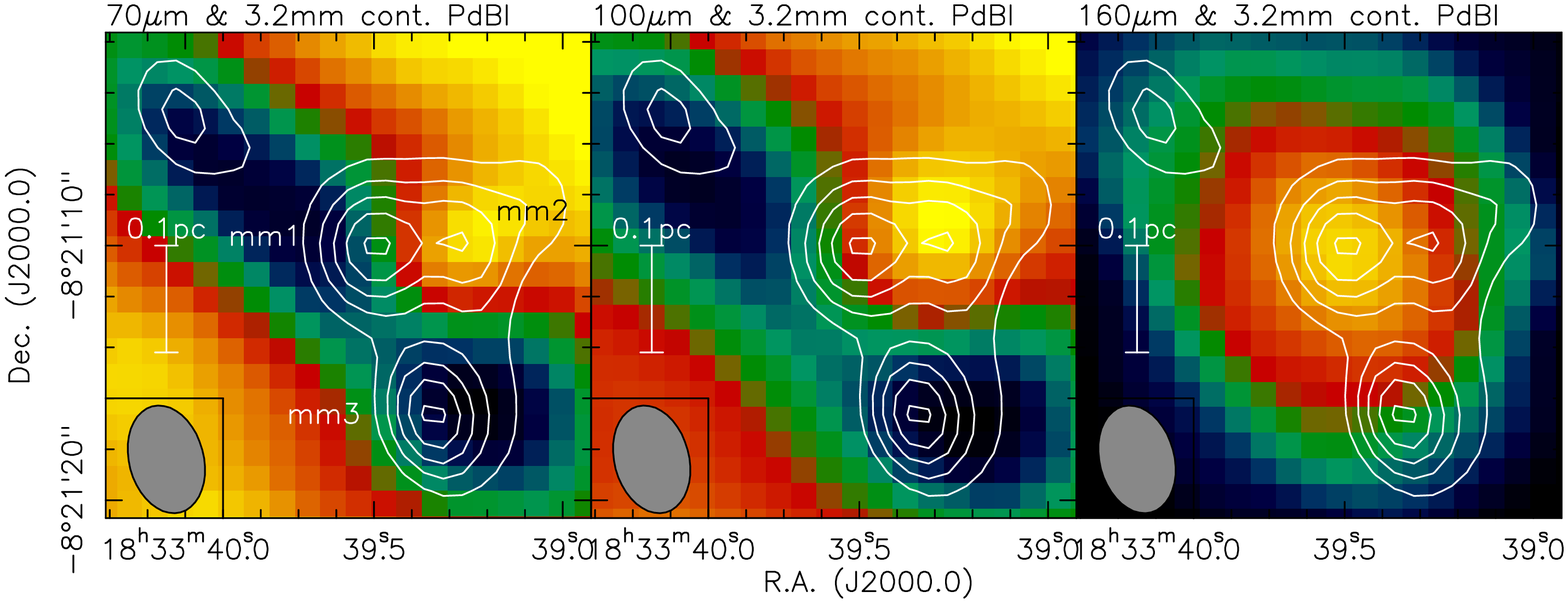}
\caption{Re-calibrated Herschel PACS images at 70, 100 and 160\,$\mu$m
  (from left to right). The scaling in the three panels is from -3.8
  to 4.0, from 5.0 to 17.6, and from 33 to 45\,mJy\,pixel$^{-1}$,
  respectively. The 3.2\,mm continuum contours are from 3$\sigma$ and
  continuing in 2$\sigma$ steps with a $1\sigma$ value of
  0.08\,mJy\,beam$^{-1}$. A scale-bar and the 3.2\,mm synthesized beam
  are shown in each panel, and the three main sources are labeled in
  the bottom-left panel. The 70, 100, and 160\,$\mu$m PSFs are
  $5.6''$, $6.8''$ and $11.3''$, respectively.}
\label{herschel}
\end{figure*}

Figure \ref{continuum} presents a compilation of the 70\,$\mu$m and
1.2\,mm single-dish data \citep{ragan2012b,beuther2002a}, the previous
3.2\,mm PdBI continuum observations \citep{beuther2013a} as well as
our new 1.07\,mm PdBI continuum maps. The three datasets cover a broad
range of resolution elements from $11''$ in the MAMBO 1.2\,mm data, to
$4.3''\times 3.0''$ in the 3.2\,mm PdBI data, to $0.6''\times 0.49''$
in our new 1.07\,mm observations. At the given distance of 4.9\,kpc,
this corresponds to linear resolution elements of $\sim 54000$\,AU,
$\sim 18000$\,AU and $\sim 2700$\,AU, respectively. With the different
resolution elements, we can clearly identify hierarchical
fragmentation on the different scales of our observations: the
large-scale single-dish gas clump with a total mass of $\sim
800$\,M$_{\odot}$ fragments into 4 cores with masses between 18 and
36\,M$_{\odot}$ \citep{beuther2013a}, and these cores again fragment
into smaller sub-structures in our new 1.07\,mm continuum data.  For
our following analysis, we only consider 1.07\,mm sources that are
detected at a $>4\sigma$ level. We can estimate the masses and column
densities of these smallest-scale structures with similar assumptions
as taken in \citet{beuther2013a}, namely optically thin dust emission
at a low temperature of 15\,K (see also sect.~\ref{pacs}), with a
gas-to-dust ratio of 150 \citep{draine2011} and dust properties
discussed in \citet{ossenkopf1994} for thin ice mantles at densities
of $10^5$\,cm$^{−3}$ ($\kappa_{1.07{\rm mm}}\sim
0.95$\,cm$^2$\,g$^{-1}$). Main uncertainties for the mass and column
density estimates are the applied dust model and the assumed
temperature. Based on this, we estimate an accuracy within a factor 2
for these parameters. This way, we estimate the masses of the
smallest-scale substructures to values between 2.2 and
19.3\,M$_{\odot}$ (Table \ref{masses}). The difference of the sum of
the core masses compared to the large-scale single-dish mass of $\sim
800$\,M$_{\odot}$ is mainly caused by the spatial filtering of the
interferometer that traces only the densest inner cores and not the
envelope anymore. The peak column densities are very high in the
regime of $10^{24}$\,cm$^{-2}$, corresponding to visual extinctions of
$\sim 1000$\,mag.

\begin{table}[htb]
\caption{Fluxes, masses and column densities}
\begin{tabular}{lrrrr}
\hline \hline
 & $S_{\rm{int}}$ & $S_{\rm{peak}}$ & $M$ & $N_{\rm{H_2}}$ \\
 & (mJy) & $\left(\frac{\rm{mJy}}{\rm{beam}}\right)$ & (M$_{\odot}$) & ($10^{24}$cm$^{-2}$) \\
\hline
mm1 & 12.0$^1$ & & 9.7$^1$ & \\
mm1a& 8.2 & 4.1 & 6.6 & 1.1 \\
mm1b& 3.8 & 2.9 & 3.1 & 0.8 \\
mm2 & 17.0$^2$ & & 13.7$^2$ & \\
mm2a& 4.5 & 3.8 & 3.6 & 1.0 \\
mm2b& 4.7 & 3.8 & 3.8  & 1.0 \\
mm2c& 3.3 & 3.3 & 2.7  & 0.9 \\
mm3a& 24.0 & 14.0 & 19.3 & 3.8 \\
mm3b& 2.7 & 2.7 & 2.2  & 0.7 \\
mm3c& 3.5 & 3.5 & 2.8  & 0.9 \\
\hline \hline
\end{tabular}
~\\
{\footnotesize Uncertainties for  $S_{\rm{int}}$ and $S_{\rm{peak}}$ are $\approx$15\% (\S \ref{obs}). Uncertainties for$M$ and $N_{\rm{H_2}}$ are approximately a factor 2 (\S \ref{cont}).\\ $^1$ Integrated over mm1 \& mm1a\\ $^2$ Integrated over mm2a, mm2b \& mm2c}
\label{masses}
\end{table}

\begin{table}[htb]
\caption{Projected nearest neighbor separations}
\begin{tabular}{lrr}
\hline \hline
 & ('') & (AU) \\
\hline 
3mm data \\
mm1--mm2 & 2.9 & 14200 \\
mm1--mm3 & 7.1 & 34900 \\
\hline
1mm data \\
mm1a--mm1b  & 0.9 & 4400 \\
mm2a--mm2b & 0.5 & 2600 \\
mm2a--mm2c & 0.7 & 3500 \\
mm2b--mm2c & 1.1 & 5500 \\
mm3a--mm3b  & 1.5 & 7600 \\
mm3a--mm3c  & 2.5 & 12000 \\
mm3b--mm3c & 2.4 & 11900 \\  
\hline \hline
\end{tabular}
~\\
{\footnotesize Uncertainties are $\approx$0.1$''$ or $\approx$490\,AU.}
\label{separations}
\end{table}

In addition to the masses and column densities, the data allow us to
estimate the fragmentation properties of the sub-sources, and we find
projected separations between the newly identified structures between
2600 and 12000\,AU (Table \ref{separations}). Since the absolute peak
positions can be determined to high accuracy ($\frac{\phi}{S/N}$ with
the resolution $\phi$ and the signal-to-noise ratio $S/N$, e.g.,
\citealt{reid1988}), and the intereferometric positional accuracy
depends mainly on the position of the used quasar and its gain
calibration solution (within $0.1''$), the projected separation is
estimated to be accurate within $0.1''$ or 490\,AU. A fragmentation
discussion is presented in section \ref{fragmentation}.


\subsection{Far-infrared continuum emission}
\label{pacs}

Figure \ref{herschel} presents a comparison of the Herschel PACS 70,
100 and 160\,$\mu$m data with the 3.2\,mm continuum data from the
PdBI.  At 70\,$\mu$m, large parts of the central region containing the
millimeter cores still appear as a dark extinction silhouette in front
of the extended 70\,$\mu$m emission of the hosting molecular cloud
(Figures \ref{continuum} \& \ref{herschel}). Although the extinction
contrast decreases at 100\,$\mu$m, we still perceive the central parts
as an infrared dark cloud (IRDC). At 160\,$\mu$m, the extended
emission in the surroundings is not so dominating anymore, and
emission structures emerge from the inside of the IRDC.

At the north-western border of the IRDC silhouette, we notice a faint
emission point source at 70\,$\mu$m which also persists at 100~$\mu$m.
In the previous data reduction products we used in \citet{ragan2012b}
and \citet{beuther2013b}, this source was not that apparent. We
attribute the differences to the use of the newest Scanamorphos
algorithms and the inclusion of the gyro-correction information
(Sect.~\ref{Subsec:Herschel-obs}) in the current data reduction. In
order to determine accurate positions, we referenced the PACS maps
with the Spitzer/MIPSGAL \citep{Carey2009} 24\,$\mu$m data. We did not
use individual MIPSGAL maps for reference, but relied on the recently
released MIPSGAL 24\,$\mu$m point source catalogue
\citep{guthermuth2015}. Corresponding positions in the PACS maps were
determined by employing PSF photometry using IDL/Starfinder
\citep{diolaiti2000}. The necessary adjustments were shifts on the
order of 1.1--1.6$''$. For the previous Herschel data products (with
uncorrected pointing) we had to use a shift of almost $4''$
\citep{beuther2013b}. This supports predictions that the absolute
pointing error of Herschel data can be brought down to a level of
$\sim$0\farcs9 1$\sigma$ if all the recent corrections for the
pointing product are applied \citep{sanchez-portal2014}. This
70\,$\mu$m point source is thus located very close to mm2.
Interestingly, the compact emission seen at 160\,$\mu$m is shifted
towards the center of the IRDC and basically coincides with the mm1
location (Fig.~\ref{herschel}). The uncertainties of the peak
positions are on the order of 1$''$ for 70 and 100\,$\mu$m, due to the
position bootstrapping involved. For the 160\,$\mu$m astrometry, the
uncertainty might even be up to $2''$, since at this wavelength it is
hard to find real point sources, and due to extended emission and
large beam sizes, the effective peak position may experience subtle
displacements (cf.~the comparison of MIPS-70 and PACS-70\,$\mu$m peaks
for the EPoS source UYSO1 reported in \citealt{linz2010}). Still, the
formal distance between the 70\,$\mu$m point source and the
160\,$\mu$m compact emission peak is more than 4$''$. Although this is
less than the beam of the 70\,$\mu$m image, as mentioned in \S
\ref{cont}, source peak positions can be determined to much higher
accuracy than the nominal spatial resolution $\phi$ (down to
$\frac{\phi}{S/N}$). Hence, we think that there is a real shift in
peak positions when going to longer wavelengths. This may be
explainable by the combined action of two effects. First, the
160\,$\mu$m data trace colder dust expected in the deeper interior of
the IRDC, compared to the 70\,$\mu$m wavelength range. Second, the
160\,$\mu$m beam is around 11\farcs3. It thus convolves emission from
larger areas, and therefore may comprise emission from mm1
(dominating) and mm2 (minor contribution), while the finer 70\,$\mu$m
beam of 5\farcs6 can still distinguish between mm2 (associated with
70\,$\mu$m emission at its north-west side) and mm1 (without
noticeable 70\,$\mu$m emission).

\begin{figure*}[htb]
\includegraphics[width=0.99\textwidth]{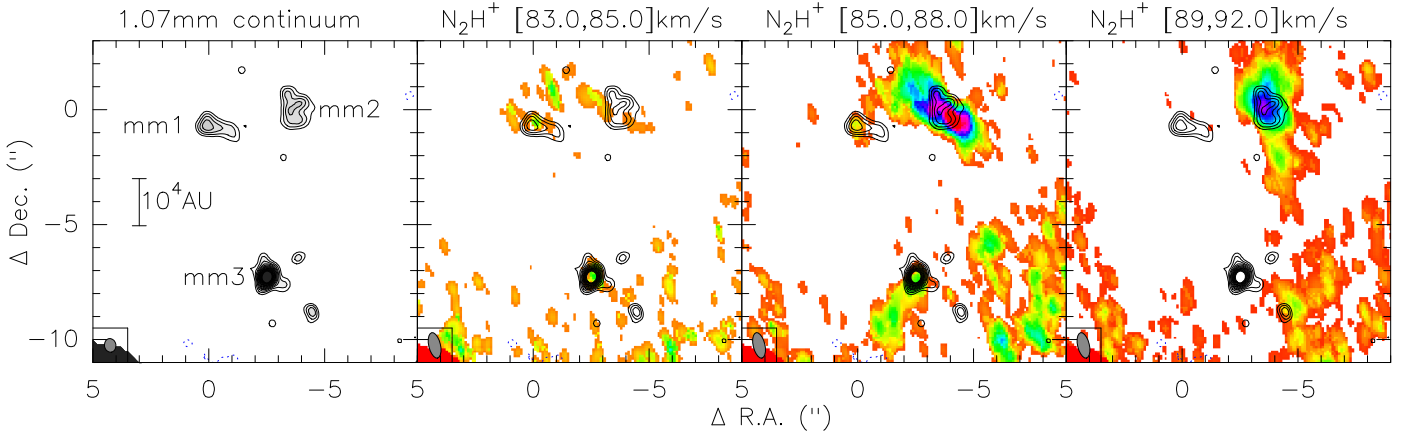}
\caption{N$_2$H$^+(3-2)$ and 1.07\,mm continuum data. The left panel
  and the contours in the three other panels always show the 1.07\,mm
  continuum data starting from 3$\sigma$ and continuing in 1$\sigma$
  steps with a $1\sigma$ value of 0.6\,mJy\,beam$^{-1}$. The
  color-scale in the 2nd, 3rd and 4th panels present the
  N$_2$H$^+(3-2)$ integrated emission over the velocity regimes marked
  above each panel. The scaling ranges in panel 2 to 4 are [0,0.15],
  [0,0.25] and [0,0.4]\,Jy\,km\,s$^{-1}$, respectively. The
  synthesized beam of the continuum and line data is shown each time
  in the bottom-left. A scale-bar can be found in the left panel.}
\label{n2h+}
\end{figure*}

\begin{table}[htb]
\label{mm2_fir}
\caption{Far-infrared fluxes of mm2}
\begin{tabular}{lr}
\hline \hline
$\lambda$ & $S$ \\
($\mu$m) & (mJy) \\
\hline
70 & 90 \\
100 & 357 \\
160$^1$ & $\leq$339 \\
3230$^2$ & 0.9 \\
\hline \hline
\end{tabular}
~\\
Flux uncertainties at 70 and 100\,$\mu$m are $\sim$10\%.\\
$^1$ Upper limit from the residual flux toward mm2 after subtracting the point source centered on mm1.\\
$^2$ From \citet{beuther2013b} with an accuracy of $\sim$15\%.
\end{table}

Hence, the PACS data show tentatively that at least object mm2
may not be totally starless anymore, but that star formation processes
might have begun at its north-western border. Whether the 70\,$\mu$m
point source is already a recently formed protostar, or just a
temperature enhancement created by various processes triggered from
within mm2, is not easy to answer with the presently available data.
At 160\,$\mu$m, we eventually see the cold dust emission from the bulk
of the core material in the mm1/2 region.

The Herschel data can also be used to get a rough estimate of the
luminosity of the mm2 core. In the 70 and 100\,$\mu$m band, we were
able to derive the far-infrared fluxes toward mm2 from fitting the
point-spread-function (PSF) to the compact emission source visible in
Fig.~\ref{herschel}. The uncertainties for the 70 and 100\,$\mu$m flux
measurements are $\approx$10\%, including $\sim$5\% calibration
uncertainty and another $\sim$5\% from the PSF photometry. At
160\,$\mu$m, this is more difficult because the emission peaks at
mm1. Therefore, in this band we can only derive an upper limit for mm2
by fitting a point source to mm1, and afterwards deriving the mm2
upper limit from the residual image. The 3\,mm flux density was taken
from \citet{beuther2013b}. Fitting a spectral energy distribution to
these four data points assuming a modified blackbody function
accounting for the wavelength-dependent emissivity of the dust, we get
an estimate of the luminosity $L$ and cold dust temperature
$T_{\rm{dust}}$. For mm2, this results in $L\sim
16_{-8}^{+14}$\,L$_{\odot}$ and $T_{\rm{dust}}\sim
15_{-0.5}^{+2.0}$\,K. The error budget in $L$ and $T_{\rm{dust}}$
includes the uncertainties of the flux calibration as well as the
selected dust model. While the dust temperature reflects the overall
cold nature of this region, the comparably low internal luminosity of
mm2 around 16\,L$_{\odot}$ shows that this region with a large mass
reservoir of $\sim$800\,M$_{\odot}$ (section \ref{cont}) still has
formed no high-mass star yet. It should also be noted that the low
luminosity cannot be explained well with accretion processes on
compact protostars because even then the accretion luminosity is
expected to be higher (e.g., \citealt{krumholz2006b}). Hence, the
observed luminosity may likely stem from accretion processes on larger
surfaces, re-enforcing the early evolutionary stage at the onset of
star formation with active collapse processes.

Looking closer at mm3, there is no emission at 70 and 100\,$\mu$m, and
the 160\,$\mu$m image exhibits only a very weak extension from mm1 in
the direction of mm3. This is in stark contrast to the 3.2\, and
1.07\,mm data which exhibit almost the same fluxes for mm1 and
mm3. This difference is most likely due to even lower temperatures
within the mm3 region compared to mm1.

\subsection{Spectral line emission}
\label{spec_lines}

\begin{figure}[htb]
\includegraphics[width=0.49\textwidth]{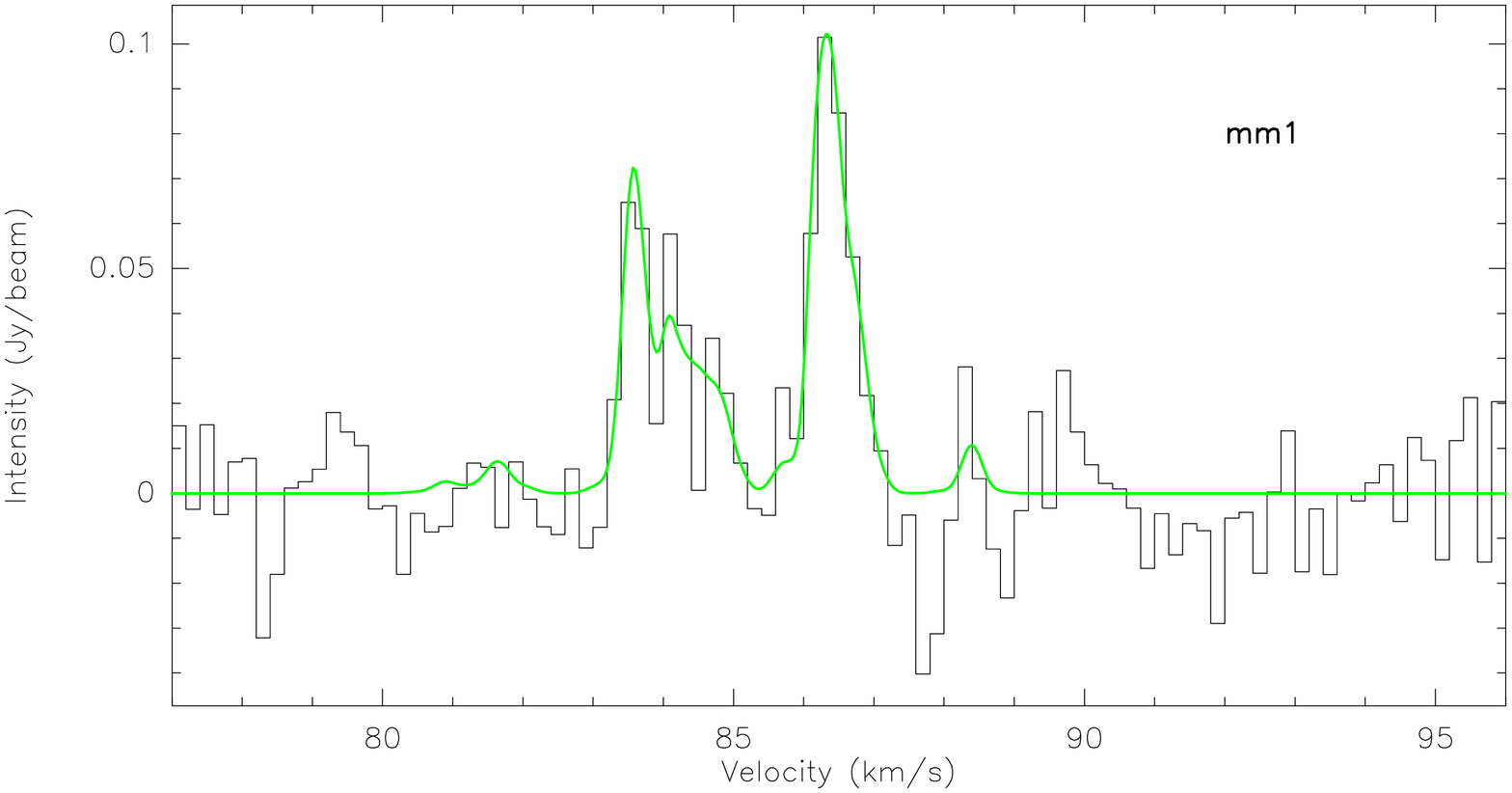}
\includegraphics[width=0.49\textwidth]{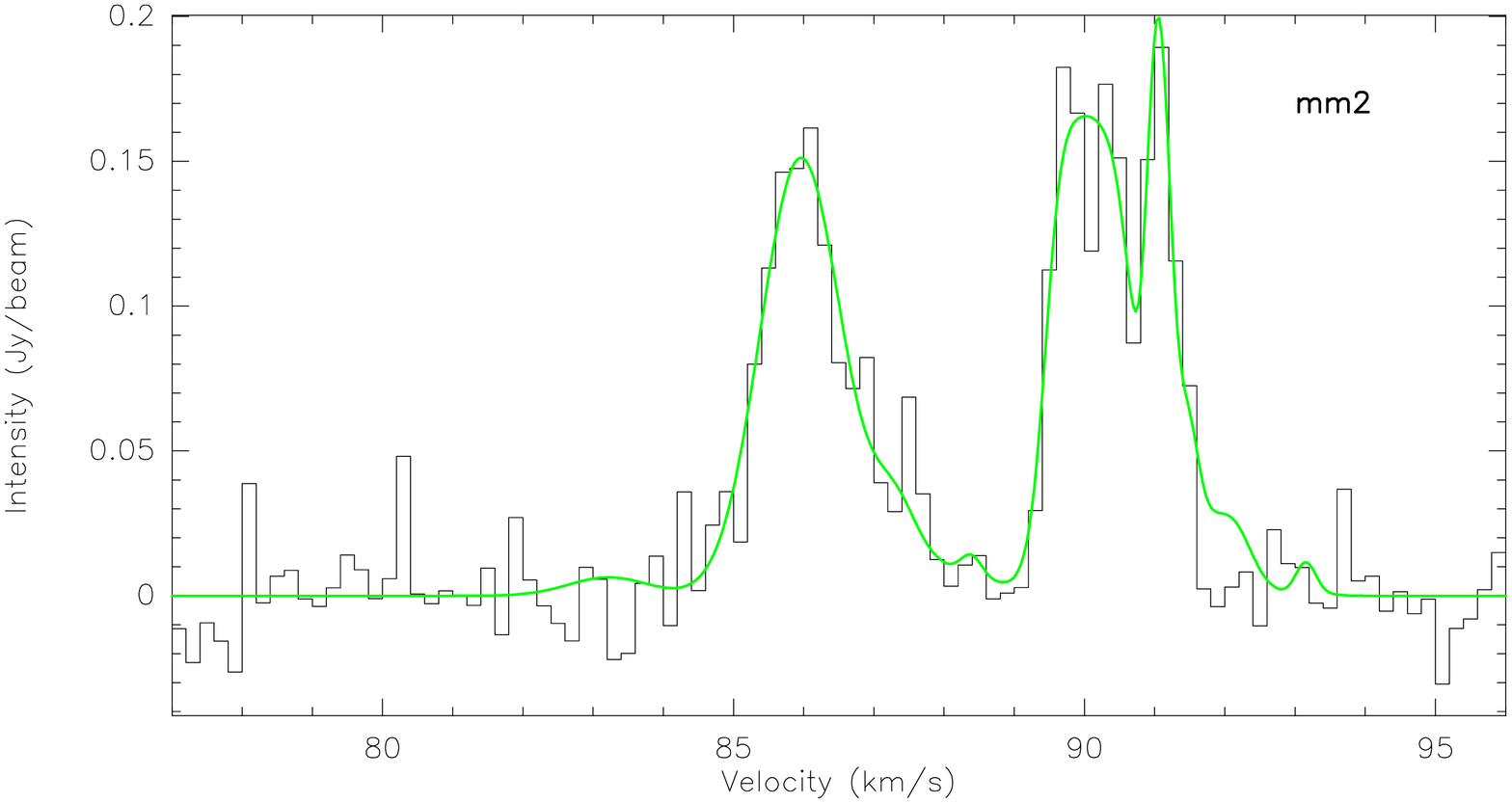}
\includegraphics[width=0.49\textwidth]{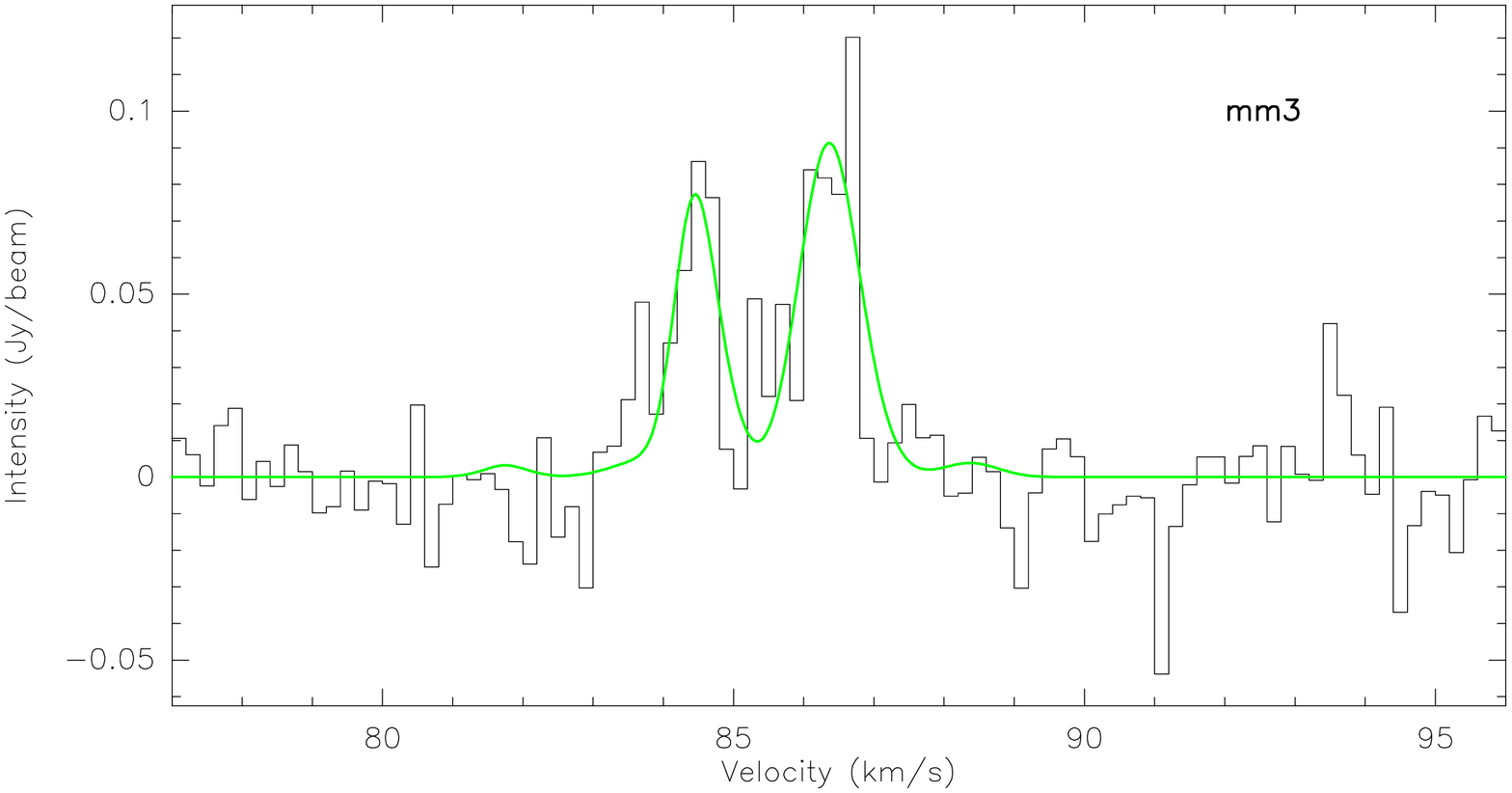}
\includegraphics[width=0.49\textwidth]{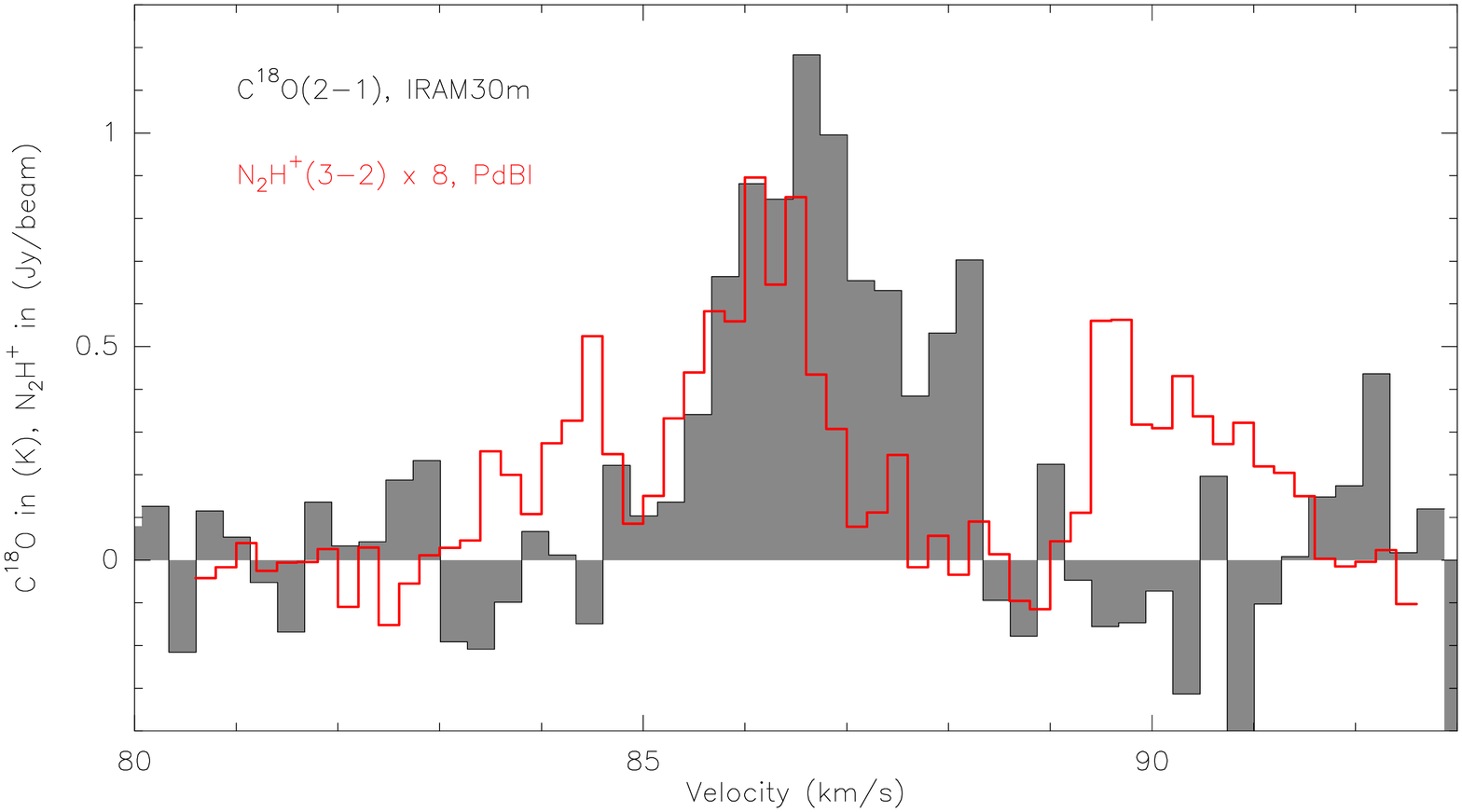}
\caption{N$_2$H$^+(3-2)$ spectra toward the three main peaks mm1, mm2
  and mm3 with the multiple component hyperfine-structure fits shown in
  green. The fit parameters are presented in Table \ref{fits}. The
  bottom spectrum shows a comparison between a single-dish
  C$^{18}$O(2--1) spectrum extracted toward the main peak in
  grey-scale, whereas the red spectrum is an average of the three
  above N$_2$H$^+(3-2)$ spectra (multiplied by 8). The units in the
  bottom spectra are different for both as marked in the side-label.}
\label{spec}
\end{figure}

Figure \ref{n2h+} presents several N$_2$H$^+$(3--2) emission maps
integrated each time over different velocity regimes. The velocity
regimes are selected based on the integrated N$_2$H$^+$(3--2) spectrum
presented in the bottom panel of Fig.~\ref{spec}. We divide the
integrations into a low-velocity part between 83 and 85\,km\,s$^{-1}$,
an intermediate-velocity regime between 85 and 88\,km\,s$^{-1}$ and a
high-velocity part from 89 to 92\,km\,s$^{-1}$. The immediate result
of that is that the different velocities trace different parts of the
dense gas. While the low-velocity component is mainly associated with
the sub-sources mm1 and mm3, the intermediate-velocity component emits
toward all three mm cores, and finally the high-velocity component
emits only toward mm2. Although the N$_2$H$^+$(3--2) line has
hyperfine-structure as well, the satellite lines are comparably weak,
and the velocity structure is mainly caused by real structure imposed
on the main hyperfine component. Nevertheless, we also fit the spectra
taking into account the full hyperfine structure of the line (see
below). Independent of that, these integrated emission images already
show that this high-mass starless clump is far from being a
kinematically homogeneous and potentially calm structure, but in
contrast to that, we see a kinematically complex region that may be at
the verge of collapse (see also \citealt{ragan2015}).

To investigate the kinematics in more depth, we extracted the spectra
toward the three main mm cores mm1, mm2 and mm3, which are shown in
Figure \ref{spec}. Again, the multiple velocity components are
apparent. To quantify the peak velocities $v_{\rm{peak}}$ and
full-width half maximum $\Delta v$, we fitted the full
N$_2$H$^+$(3--2) hyperfine structure to these lines with multiple
components. Note that, in contrast to the 1--0 transition, for the
3--2 transition of N$_2$H$^+$ the spectrum is much more strongly
dominated by the central line component. From the 29 hyperfine
structure components more than 60\% of the relative intensities are
within the central 6 components separated by only
$\sim$0.06\,km\,s$^{-1}$ whereas the remaining 23 components share the
rest of the emission. The resulting fits are shown in Table
\ref{fits}. One should keep in mind that the selection of the number
of fitted velocity components is partly ambiguous. For example, for
mm2 we fitted in total three components but one could also have fitted
the high-velocity feature in mm2 with three components alone and have
a fourth component around the intermediate-velocity gas. Therefore,
these fits do not claim to be the final answer, but they nevertheless
adequately represent the current data. While the multiple components
toward individual peaks are interesting in themselves, also the
derived line width are important. They range in
full-width-half-maximum (FWHM) between 0.3 and 1.3\,km\,s$^{-1}$ where
the 0.3\,km\,s$^{-1}$ have to be considered as an upper limit because
with our velocity resolution of 0.2\,km\,s$^{-1}$ no narrower lines
can be resolved. The broad end of the distribution is most likely also
an upper limit, however this time caused by unresolved underlying
spectral multiplicity in the lines. With the thermal line width at
15\,K of $\sim$0.15\,km\,s$^{-1}$, these new data now resolve the
spectra into almost thermal lines, just with multiple components.

For comparison, Figure \ref{spec} also shows a single-dish
C$^{18}$O(2--1) spectrum at $11''$ resolution (Ragan et
al.~priv.~comm.) as well as the PdBI N$_2$H$^+$(3--2) spectrum
integrated over all three mm continuum cores. While the C$^{18}$O
spectrum easily traces the main intermediate-velocity component,
however spectrally not well resolved in the sub-components (a Gaussian
two-component fit to the C$^{18}$O(2--1) spectrum results in $\Delta
v$ values of $\sim$1.7 and $\sim$0.3\,km\,s$^{-1}$, a 1-component fit
has $\Delta v\sim 2.0$\,km\,s$^{-1}$), the high- and low-velocity
components are not seen at all in the C$^{18}$O(2--1) spectrum. The
reason for that is likely at least twofold and lies within the lower
spatial resolution of the single-dish data as well as the about two
orders of magnitude lower critical density of C$^{18}$O(2--1) compared
to N$_2$H$^+$(3--2) ($\sim 10^4$ versus $\sim 10^6$\,cm$^{-3}$). With
the new high-resolution observations of this dense gas tracer, we are
able to really dissect the densest portions of this very young
high-mass star-forming region (see section \ref{kinematics}).

\begin{table}[htb]
\caption{Hyperfine structure fits toward the mm peaks}
\label{fits}
\begin{tabular}{lrr}
\hline 
\hline
  & $v_{\rm{peak}}$ & $\Delta v$ \\
  & (km\,s$^{-1}$) &  (km\,s$^{-1}$)\\
\hline
mm1 & 83.5 & 0.3 \\
mm1 & 84.3 & 0.3 \\
mm1 & 86.3 & 0.3 \\
\hline
mm2 & 85.8 & 1.3 \\
mm2 & 89.9 & 0.7 \\
mm2 & 91.0 & 0.3 \\
\hline 
mm3 & 84.4 & 0.6 \\
mm3 & 86.2 & 0.9 \\
\hline 
\hline
\end{tabular}
\end{table}

\section{Discussion}

We now have the ability to track the hierarchical fragmentation and
global collapse of starless gas clumps capturing smaller and smaller
scales.

\subsection{Fragmentation}
\label{fragmentation}

Following \citet{beuther2013a}, we can estimate the average densities
$\rho$ for the large-scale single-dish data as well as the previous
intermediate-scale 3\,mm continuum data, corresponding to the
bottom-left and bottom-middle panels in Figure \ref{continuum}.  The
estimated average densities $\rho$ derived from the two datasets are
$2.5\times 10^5$\,cm$^{-3}$ and $1.5\times 10^6$\,cm$^{-3}$,
respectively. Assuming furthermore average temperatures of 15\,K,
\citet{beuther2013a} estimated the Jeans length $\lambda_{\rm Jeans}$
and Jeans mass $M_{\rm Jeans}$ that predict the expected fragment
properties of the corresponding smaller scales in the framework of
this isothermal gravitational Jeans fragmentation picture. The
$\lambda_{\rm Jeans}$ and $M_{\rm Jeans}$ based on the large-scale
single-dish data are 10000\,AU and 0.37\,M$_{\odot}$, whereas the
corresponding $\lambda_{\rm Jeans}$ and $M_{\rm Jeans}$ derived from
the intermediate-scale 3\,mm PdBI data are 4000\,AU and
0.15\,$M_{\odot}$ \citep{beuther2013a}.

If we now compare the predicted length and mass scales with the
corresponding observed values on the different scales, we find
correspondences as well as differences. Regarding the Jeans length and
the observed projected separations, the predicted Jeans length from
the single-dish data of $\sim$10000\,AU is slightly smaller than the
core separation in the PdBI 3.2\,mm data (Table \ref{separations}),
but the values are consistent within a factor of a few. Going to
smaller scales, the predicted Jeans length from the 3.2\,mm data of
$\sim$4000\,AU is roughly consistent with the projected separations we
find again at the smaller scales of the 1.07\,mm continuum data (Table
\ref{separations}). Hence, from a pure length-scale argument, the
observations of IRDC\,18310-4 would be roughly consistent with
classical thermal Jeans fragmentation. However, our observations
reveal only projected separations, and the real ones can be roughly
up to a factor 2 larger.

Does this picture also hold for the Jeans masses? In fact, it does not
because the predicted Jeans masses on the different scales of 0.37 and
0.15\,$M_{\odot}$, respectively, are significantly lower than what is
found in the 3.2\,mm data (masses for mm1 to mm3 between 18 and
36\,$M_{\odot}$, \citealt{beuther2013a}) as well as in our new
1.07\,mm observations (masses between 2.2 and 13.7\,M$_{\odot}$, Table
\ref{masses}). Hence, while the length scales are roughly consistent
between the classical Jeans predictions and the data, the masses
deviate by up to two orders of magnitude. We note that these observed
masses are even only lower limits because large fractions of the gas
are filtered out.

Similar discrepancies were recently reported by \citet{wang2014} in
their fragmentation study of parts of the ``Snake'' filament. They
also found projected separations in their data that correspond
reasonably well to the estimated Jeans length, whereas the fragment
masses exceeded the Jeans masses by similar margins as in our
data. They argued that the typical thermal Jeans mass is calculated
with the thermal sound speed depending on the temperature of the
gas. However, \citet{wang2014} evaluate a turbulent Jeans mass using
the turbulent velocity dispersion of the gas instead of the velocity
dispersion based on the thermal sound speed. Since the Jeans mass
depends on the velocity dispersion to the third power, even velocity
dispersion increases of a factor of a few allow to shift the turbulent
Jeans masses by more than an order of magnitude in their regime of
observed masses. Hence, \citet{wang2014} argue that turbulent Jeans
fragmentation may explain the observed properties. A first-order
inconsistency could arise in this picture when considering the lengths
scales. In the turbulent Jeans scenario, also the turbulent Jeans
length has to be adapted according to the turbulent velocity
dispersion. Although the velocity dispersion only enters the equation
for the Jeans length linearly, nevertheless, it increases the
turbulent Jeans length by a factor of a few. And in that picture, the
projected separations of their cores would be a factor of a few
smaller than the predicted turbulent Jeans length. However, taking
into account potential projection effects, this difference may be less
severe.

This turbulent Jeans fragmentation picture becomes questionable for
IRDC\,18310-4 if one considers that toward one of our most massive
sub-condensations mm1a with 6.6\,M$_{\odot}$ the measured line widths
only shows upper limits of 0.3\,km\,s$^{-1}$.  Hence, no significant
turbulent contribution to the line-width is observed here.

A different approach to solve the discrepancies between the classical
Jeans fragmentation and our data is to investigate the initial
conditions, in particular the initial density structure. While the
Jeans fragmentation analysis starts with uniform and infinite gas
structures (e.g., \citealt{stahler2005}), the derived density structures
of the observed gas clumps exhibit much steeper density profiles
(e.g., \citealt{beuther2002a,mueller2002,hatchell2003}).
\citet{girichidis2011} analyzed the different fragmentation properties
during star formation in numerical simulations varying the turbulent
velocity fields as well as the density profiles from uniform to
Bonnor-Ebert sphere and density power-law profiles $\rho\propto
r^{-p}$ with indices $p$ of 1.5 and 2. As expected, the steeper the
initial density profile is, the less fragmentation they find and the
more massive the final fragment masses are. In the extreme cases, the
difference in masses rises by up to a factor $\sim$20. While they do
not explicitly give the fragment separations, the fact that they find
less fragments, indicates that the fragment separation should also
increase.

To summarizes these effects, most likely, the observed fragmentation
properties in IRDC\,18310-4, as well as other sources in the
literature, could be explained by turbulence-suppressed fragmentation of
gas clumps with non-uniform density structures. Magnetic fields can
be another agent in inhibiting fragmentation (e.g.,
\citealt{commercon2011,pillai2015}).

\subsection{Kinematic properties}
\label{kinematics}

As outlined in section \ref{spec_lines}, our new high-spatial and
high-spectral resolution observations of the dense gas tracer
N$_2$H$^+$(3--2) resolve the spectral lines from the several cores
into multiple spectral components where the individual components are
only slightly above the thermal line width. Although the data are not
good enough yet to actually resolve coherent thermally dominated
structures like those found for example in B5
\citep{pineda2010b,pineda2015}, we are reaching a regime for the
individual sub-cores within a high-mass star-forming region that is
not that far apart from coherence.

However, in addition to the narrow line widths found for individual
components, the omnipresence of multiple components toward each core
implies additional highly dynamic processes within the overall
high-mass star-forming region. The most straightforward interpretation
of these multiple velocity components is within the picture of a
globally collapsing gas clump where the different velocity components
trace separate individual gas parcels that fall toward the
gravitational center of the whole gas clump. The spectral line
N$_2$H$^+$(1--0) signatures of such a global collapse were simulated
by \citet{smith2013}, and they find similar spectra with multiple
components along individual line of sights as seen in our
IRDC\,18310-4 data.

The C$^{18}$O(2--1) data allow us also to derive a rough estimate of
the virial mass. Following \citet{maclaren1988} assuming a density
profile $\rho\propto 1/r$, a radius of the clump of 0.25\,pc
(Fig.~\ref{continuum}) as well as the 1-component fit $\Delta v\sim
2.0$\,km\,s$^{-1}$ to the C$^{18}$O(2--1) spectrum, the approximate
virial mass is $\sim$190\,M$_{\odot}$, about a factor 4 lower than the
gas mass of 800\,M$_{\odot}$ derived from the dust continuum data
\citep{beuther2013a}\footnote{Density profiles steeper than $1/r$
  would result in even lower virial mass estimates.}.

In the framework of a globally collapsing gas clump, one can use the
observed spectral velocity differences also for a simple collapse time
estimate. Assuming that the difference between the velocity peaks is
due to cores sitting at different points in a globally collapsing
region, we get converging velocity differences along the lines of
sight of 2.8\,km\,s$^{-1}$ in mm1, 5.2\,km\,s$^{-1}$ in mm2, and
1.8\,km\,$^{-1}$s in mm3. Ignoring for this estimate the plane of the
sky velocity component that we do not know, these velocity gradients
would bring together fragments 10000\,AU apart in only $\sim 1\times
10^4$ to $\sim 2.6\times 10^4$\,yrs. The collapse of the clump may
also cause the fragments to merge or increase in density, which could
help explain the high fragment masses in the previous section. For
example, \citet{smith2009a} found, using synthetic interferometry
observations of simulated gas clumps, that the number of fragments
decreased and their mean column density increased as the clumps
collapsed. This could efficiently increase the mass of the fragments.

Combining the multiple components with the individual narrow line
widths as well as the low virial mass, we interprete the kinematic
data in this region as most likely caused by a dynamical collapse of a
large-scale gas clump that caused multiple velocity components along
the line of sight, and where at the same time the individual infalling
gas structures have very low levels of internal turbulence.

\section{Conclusions}

Resolving the mm continuum and N$_2$H$^+$(3--2) emission at
sub-arcsecond resolution (linear scales down to $\sim$2500\,AU)
toward the pristine high-mass starless gas clump IRDC\,18310-4, we
reveal the fragmentation and kinematic properties of the dense gas at
the onset of massive star formation. Zooming through different size
scales from single-dish data to intermediate- and high-angular
resolution PdBI observations, the resolved entities always fragment
hierarchically into smaller sub-structures at the higher spatial
resolution. While the fragment separations are still in approximate
agreement with thermal Jeans fragmentation, the observed core masses
are orders of magnitude larger than estimated Jeans masses at the
given densities and temperatures.  Hence, additional processes have to
be in place. However, taking into account non-uniform density
structures as well as initial turbulent gas properties, observed core
masses and projected separations are consistent with cloud formation
models.

While most sub-cores are (far-)infrared dark even at
70\,$\mu$m, the re-reduced Herschel data reveal weak 70\,$\mu$m
emission toward core mm2 with a comparably low luminosity of
only $\sim$16\,L$_{\odot}$. Since such a low luminosity can be caused
neither by an internal high-mass star nor by strong accretion onto a
typical embedded protostar, this re-enforces the youth and early
evolutionary stage of the region.

The spectral line data reveal multiple velocity components with
comparably small width (still above thermal) toward the individual
sub-cores. Relating these data to cloud collapse simulations, they are
agreeing with globally collapsing gas clumps where several gas
parcels along the line of sight are revealed as individual spectral
features. The narrow line widths in the regime 0.3 to 1\,km\,s$^{-1}$
indicate that even during the dynamical global collapse, individual
sub-parcels of gas can have very low internal levels of turbulence.





\end{document}